\documentclass[10pt,twocolumn,floatfix,superscriptaddress,preprintnumbers,amsmath,amssymb,prl]{revtex4}
\usepackage{graphicx,bm}
\usepackage{amssymb,amsmath}
\usepackage{float}
\usepackage{epstopdf}
\usepackage[dvips]{color}
\newcommand{\TRB} {TaRh$_{2}$B$_{2}$}
\newcommand{\TC} {$T_{\mathrm{c}}$}

\begin{document}

\title{Multigap Superconductivity in Chiral Noncentrosymmetric TaRh$_{2}$B$_{2}$}

\author{D. A. Mayoh}
\email[]{d.mayoh@warwick.ac.uk}
\affiliation{Physics Department, University of Warwick, Coventry, CV4 7AL, United Kingdom}

\author{A. D. Hillier}
\affiliation{ISIS facility, STFC Rutherford Appleton Laboratory, Harwell Science and Innovation Campus, Oxfordshire OX11 0QX, United Kingdom}

\author{K. G\"{o}tze}
\affiliation{Physics Department, University of Warwick, Coventry, CV4 7AL, United Kingdom}

\author{D. McK. Paul}
\affiliation{Physics Department, University of Warwick, Coventry, CV4 7AL, United Kingdom}

\author{G. Balakrishnan}
\affiliation{Physics Department, University of Warwick, Coventry, CV4 7AL, United Kingdom}

\author{M. R. Lees}
\affiliation{Physics Department, University of Warwick, Coventry, CV4 7AL, United Kingdom}

\begin{abstract}

We report the first observation of multigap superconductivity in \TRB. We show \TRB\ is a bulk type-II superconductor with a transition temperature, $T_{\mathrm{c}} = 6.00(5)$~K. We present transverse-field muon spin relaxation data where the superconducting gap can be fit using a two-gap $\left(s+s\right)$-wave model. We also report the zero-field electronic specific heat in the superconducting state that is best described by the same $\left(s+s\right)$ model providing further evidence of multiband behavior in this superconductor. Zero-field muon spin relaxation measurements show time-reversal symmetry is preserved in the superconducting state. We demonstrate that \TRB\ has an upper critical field of $15.2(1)$~T, which is significantly higher than previously reported and exceeds the Pauli limit.
\end{abstract}

\maketitle

\textit{Introduction}:- The discovery of two noncentrosymmetric (NCS) superconductors with chiral structures, \TRB\ and NbRh$_{2}$B$_{2}$,~\cite{Carnicom} has added a new twist to an already exciting area of superconductivity research~\cite{Bauer12, Smidman17}. For all superconductors, the topology of the electronic band structure along with the underlying crystal structure, play vital roles in determining the superconducting properties of the material. 

For a simple $s$-wave, electron-phonon mediated superconductor with a single Fermi surface, the gap function has the full symmetry of the crystal. Higher angular momentum ($p$-, $d$-, and $f$-wave) coupling channels can lead to nodes in the gap and if, e.g., two equivalent gap functions are degenerate, a chiral state such as $\left(p_x+ip_y\right)$ where the gap function winds around some axis of the Fermi surface. Chiral superconductivity may also arise due to the presence of several Fermi sheets, requiring multiple gap functions to describe the superconducting state. Chiral superconductors can exhibit broken time reversal symmetry (TRS) and support exotic topologically protected edge states, surface currents, and Majorana modes~\cite{Kallin16}. Two of the most studied candidate chiral superconductors, the heavy fermion $f$-wave superconductor UPt$_{3}$, and the $p$-wave oxide superconductor Sr$_{2}$RuO$_{4}$, have multiple bands crossing the Fermi energy. More recently it has been suggested that SrPtAs, a superconductor with a locally NCS structure, exhibits a chiral $\left(d+id\right)$-wave state~\cite{Fischer14}. Both Sr$_{2}$RuO$_{4}$ and SrPtAs break time reversal symmetry~\cite{Sr2RuO4, SrPtAs}. 

In contrast to superconductors with a centrosymmetric (CS) crystal structure where parity is a good quantum number and the Cooper pairs form in a purely spin-singlet or spin-triplet state, the lack of a center inversion in a NCS superconductor means parity is is no longer a good quantum number. Antisymmetric spin-orbit coupling (ASOC) can then lead to a mixing of the singlet and triplet pair states. This mixing gives rise to interesting behavior including time-reversal symmetry breaking~\cite{LaNiC2, Re6Zra, La7Ir3} and upper critical fields that exceed the Pauli limit. NCS superconductors with a chiral structure are rare but examples do exist. For example, Li$_{2}$(Pd$_{1-x}$Pt$_x$)$_{3}$B $\left(0 \leq x\leq 1\right)$ has an antiperovskite structure, (NCS chiral space group $P4_332$). Li$_{2}$Pd$_{3}$B is an $s$-wave superconductor, while the stronger ASOC in Li$_{2}$Pt$_{3}$B enhances the triplet component, producing line nodes in the gap, as revealed by the linear temperature dependence of the magnetic penetration depth $\lambda\left(T\right)$ at low $T$~\cite{Yuan2006, Nishiyama2007}. Mo$_{3}$Al$_{2}$C, (NCS chiral space group $P4_132$), is a strongly-coupled superconductor with a \TC\ of 9~K. NMR and electronic heat capacity data, along with a pressure enhanced \TC\ suggest Mo$_{3}$Al$_{2}$C has a nodal gap with singlet-triplet mixing~\cite{Mo3Al2C, Bauer10}. 

Multigap superconductivity has been reported in several classes of superconductor including the iron pnictides~\cite{Stewart11, Khasanov15, Si16} and the borides including MgB$_2$~\cite{Bouquet01, Liu, Iavarone, Khasanov15, Zehetmayer}. There are several non-chiral NCS superconductors that exhibit multigap behavior. For example, Y$_2$C$_3$ is proposed to have a double-gap, each with $s$-wave symmetry and interband coupling. Even in CS materials, a combination of strong spin-orbit (SO) coupling and disorder can have profound effects on the superconducting properties. The $R_2$Pd$_x$S$_5$ $\left(x\leq 1\right)$ family of materials where $R=$~Nb or Ta with the CS space group $C2/m$ are particularly noteworthy in this regard~\cite{Zhang13, Lu14, Biswas15}. Strong SO coupling, along with Anderson localization resulting from a level of disorder due to Pd deficiency, leads to highly anisotropic upper critical fields that in some cases far exceed the Pauli limit (e.g., Nb$_2$Pd$_{0.81}$S$_5$ $T_{\mathrm{c}}\sim 6.6$~K and $H_{\mathrm{c2}}\left(0\right)$ along the $b$ axis of  37~T)~\cite{Zhang13}.

\TRB\ and NbRh$_{2}$B$_{2}$ are both NCS superconductors with a chiral structure and have the potential to exhibit similar exciting physics. The first report on this material~\cite{Carnicom} shows \TRB\ has a trigonal structure (space group $P3_1$), with a \TC\ of 5.8~K and an upper critical field that exceeds the Pauli limit. Here, we have used a combination of muon spectroscopy and heat capacity to probe the nature of the superconducting state of \TRB. We show that \TRB\ can be best described using a two-gap $\left(s+s\right)$-wave model. We clearly demonstrate that \TRB\ is a type II superconductor with an upper critical field, $\mu_0H_{\mathrm{c2}}=15.2(1)$~T, that is significantly higher than previous reported~\cite{Carnicom} and well above the Pauli limit.

\textit{Experimental Details}:- Polycrystalline samples of \TRB\ were prepared from the constituent elements using the synthesis method described earlier~\cite{Carnicom, SupplMat}. Powder X-ray diffraction (PXRD) performed using a Panalytical X-Pert Pro diffractometer showed the material forms with a trigonal $P3_1$ structure and lattice parameters $a = 4.6984(2)$~\AA\ and $c = 8.7338(5)$~\AA. The samples were confirmed to be single phase to within the detection limit of the technique. 

A Quantum Design Physical Property Measurement System (PPMS) with a $^3$He insert was used to measure ac resistivity $\left(\rho\right)$ and heat capacity $\left(C\right)$ from 0.5 to 300~K in fields up to 9~T. $\rho\left(T\right)$ measurements were extended to 15~T using an Oxford Instruments cryomagnet. 

Zero-field (ZF) and transverse-field (TF) muon spin relaxation/rotation ($\mu$SR) experiments were carried out using the MuSR spectrometer at the ISIS pulsed muon facility. The muons implanted into the sample rapidly thermalize, before decaying with a mean lifetime of 2.2~$\mu$s. The decay positrons are recorded by 64 scintillation detectors positioned in circular arrays around the sample. A full description of the detector geometries is given in Ref.~\cite{SLee}. Stray fields at the sample position were canceled to within 1~$\mu$T by an active compensation system. The powdered TaRh$_{2}$B$_{2}$ sample was mounted on a 99.995\% silver plate and placed in a $^3$He refrigerator, allowing data to be collected between $0.3 \leq T \leq 8.5$~K, in transverse fields up to 40~mT.

\begin{figure}[tb]
\centering
\includegraphics[width=0.8\columnwidth]{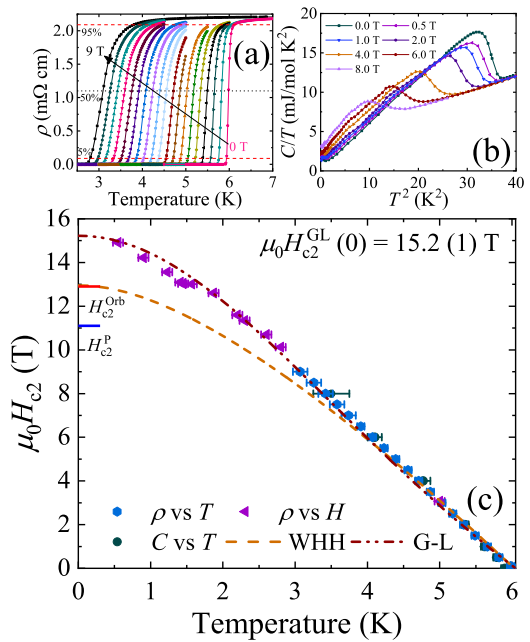}
\caption{(Color online) (a) Resistivity vs temperature for \TRB\ in fields from 0 to 9~T in 0.5~T steps. The black (red) dashed line(s) are at 50\% (95\% and 5\%) of the resistivity just above the superconducting transition. (b) $C\left(T\right)/T$ vs $T^2$ in several applied fields up to 8~T. (c) Temperature dependence of the upper critical field for \TRB. Points were extracted from the midpoint (50\%) of the resistive transition and the anomaly in $C\left(T\right)$ at the midpoint of the superconducting transition. The dashed and dashed-dotted lines show the expected $\mu_0H_{\mathrm{c2}}\left(T\right)$ from the WHH and G-L models respectively.}
\label{Resistivity}
\end{figure}

\begin{figure}[tb]
\centering
\includegraphics[width=0.8\columnwidth]{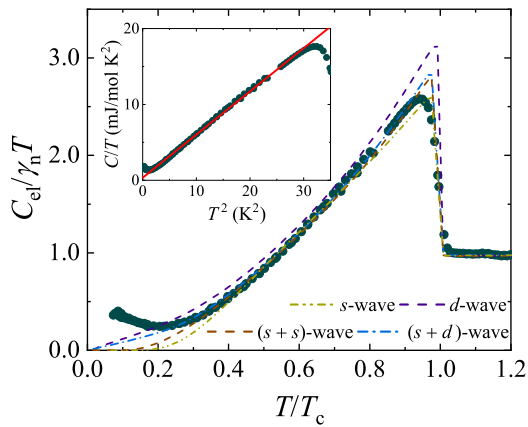}
\caption{(Color online) Zero-field normalized electronic specific heat of \TRB\ with a single-gap isotropic $s$-wave (yellow), a line nodal $d$-wave (purple), an isotropic two-gap $\left(s+s\right)$-wave (blue) and a two-gap $\left(s+d\right)$-wave model (dark red). The inset shows a linear fit to the zero-field $C\left(T\right)/T$ vs $T^{2}$ indicating a $T^{3}$ dependence to the specific heat.}
\label{SCgap}
\end{figure}

\textit{Characterization of the superconducting state}:- Heat capacity and resistivity data show \TRB\ is a bulk type-II superconductor with a superconducting transition temperature $T_{\mathrm{c}}=6.00(5)$~K. A more comprehensive set of characterization data are provided in the Supplemental Material~\cite{SupplMat}. Fig.~\ref{Resistivity}(a) shows $\rho\left(T\right)$ for \TRB\ close to \TC\ in different magnetic fields. The resistivity just above \TC\ is high ($\sim 2~\mathrm{m}\Omega$~cm) with an upturn in $\rho\left(T\right)$ suggesting a degree of localization~\cite{Lu14}. The transition broadens and shifts to lower $T$ with increasing field. An anomaly in specific heat at \TC\ confirms bulk superconductivity (see Fig.~\ref{Resistivity}(b)). $\gamma_{\mathrm{n}} = 4.29(6)$~mJ/mol K$^{2}$ and $\Delta C/\gamma_{\mathrm{n}}T_{\mathrm{c}} = 1.57(2)$ is larger than the 1.43 expected from conventional Bardeen Cooper Schrieffer (BCS) superconductors, suggesting the presence of enhanced electron-phonon (e-p) coupling. 

The critical temperatures or upper critical fields extracted from the midpoint of the resistive transition or the anomaly in $C\left(T\right)$ at the superconducting transition, are collected together in Fig~\ref{Resistivity}(c). The upper critical field $\mu_{0}H_{\mathrm{c2}}\left(T\right)$ is almost linear in $T$ at lower fields. The data deviate from the Werthamer-Helfand-Hohenberg (WHH) model~\cite{WHH} at lower $T$ and so we use a phenomenological Ginzburg-Landau expression, $H_{\mathrm{c2}}\left(T\right)=H_{\mathrm{c2}}\left(0\right)\left[1-\left(T/T_{\mathrm{c}}\right)^2\right]/\left[1+\left(T/T_{\mathrm{c}}\right)^2\right]$ to estimate $\mu_0H_{\mathrm{c2}}\left(0\right)=15.2(1)$~T, which is well above the Pauli limit of $\mu_0H^{\mathrm{P}}\left[\mathrm{T}\right]=1.85\times T_{\mathrm{c}}\left[\mathrm{K}\right]=11.1(1)$~T. The temperature dependence of the electronic specific heat, $C_{\mathrm{el}}\left(T\right)$, provides important information about the nature of the superconducting gap (see Fig.~\ref{SCgap}). Immediately below \TC, $C_{\mathrm{el}}\left(T\right)$ clearly deviates from a simple $s$-wave BCS-like behavior instead exhibiting a $T^3$ dependence (see inset Fig.~\ref{SCgap}). A similar behavior is seen in other two-gap superconductors including Lu$_2$Fe$_3$Si$_5$~\cite{Nakajima08, BiswasLu2fe3Si5} and MgB$_2$~\cite{Bouquet01, Zehetmayer}. In these materials, a shoulder in $C_{\mathrm{el}}\left(T\right)$ is followed by a rapid fall at lower $T$ as the second gap opens. As seen in Fig.~\ref{SCgap} we observe a flattening in $C_{\mathrm{el}}\left(T\right)$ in \TRB\ at low-temperature due to a hyperfine contribution. The zero-field $C_{\mathrm{el}}\left(T\right)$ data from $T_{\mathrm{c}}$ to 1.5~K were fit using a single-gap isotropic $s$-wave, a line nodal $d$-wave, an isotropic two-gap $\left(s+s\right)$-wave, and an $\left(s+d\right)$-wave model~\cite{Gapfitting} with the two-gap models producing a better fit to the data (shown Fig.~\ref{SCgap}). The gap values for the $\left(s+s\right)$-wave fit are $\Delta_{0,1}/k_{\mathrm{B}}T_{\mathrm{c}} = 2.19(12)$, $\Delta_{0,2}/k_{\mathrm{B}}T_{\mathrm{c}} = 1.1(3)$ with a weighting of 4:1. The larger gap is above the BCS value of 1.76, again indicating moderate e-p coupling.

\begin{figure}[tb]
\centering
\includegraphics[width=0.8\columnwidth]{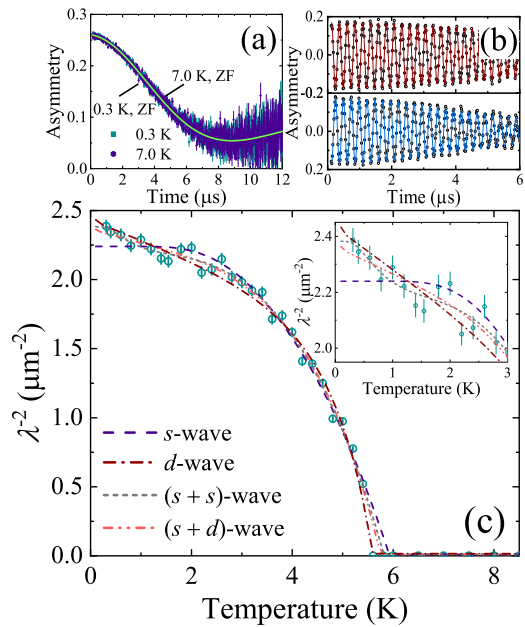}
\caption{(Color online) (a) Zero-field $\mu$SR time spectra collected at 0.3 (green) and 7.0~K (purple) show TRS is preserved in \TRB. (b) Transverse field $\mu$SR time spectra for \TRB\ collected at 7.0 (top) and 0.3~K (bottom) in an applied field of 30~mT. The solid lines are fits using Eq.~\ref{TF1}. The Gaussian decay of the oscillatory signal shows the effect of the flux-line lattice. (c) Inverse square of the London penetration depth, $\lambda^{-2}$, as a function of temperature for \TRB. The lines are fits to the data using Eqs.~\ref{London}~and~\ref{weighting} for one- and two-gap models. The inset shows the low-temperature data on an expanded scale.}
\label{ZF_TFdata}
\end{figure}

\textit{Zero-field $\mu$SR measurements}:- Zero-field (ZF) muon spin relaxation spectra collected at temperatures above (7~K) and below (0.3~K) \TC\ are shown in Fig.~\ref{ZF_TFdata}(a). In these experiments, any muons stopped in the silver sample holder give a time-independent background. There is no indication of a precessional signal ruling out the possibility of a large internal field and hence long-range magnetic order in \TRB\ down to at least 300~mK. In the absence of atomic moments, the muon spin relaxation is due to static, randomly oriented local fields associated with the nuclear moments at the muon site. The solid lines in Fig.~\ref{ZF_TFdata}(a) are fits using a Gaussian Kubo-Toyabe expression~\cite{Kubo-Toyabe}. The fact that the spectra above and below \TC\ are identical confirms time-reversal symmetry is preserved in \TRB\ in the superconducting state.
 
\textit{Transverse-field $\mu$SR measurements}:- Transverse-field $\mu$SR experiments were performed in the superconducting mixed state in applied fields of 30 and 40~mT, well above the $\mu_0H_{\mathrm{c1}}(0)\sim 4$~mT of \TRB~\cite{SupplMat}. In this geometry, the detectors in the spectrometer were grouped in eight blocks, each with a phase offset $\phi$. In order to ensure the most uniform flux-line lattice possible, the magnetic field was applied above \TC\ and the sample cooled to base temperature. The data were then collected in field-cooled-warming (FCW) mode. A typical TF-$\mu$SR precession signal for \TRB\ in 30~mT at 300~mK is shown in Fig.~\ref{ZF_TFdata}(b). Above \TC\ the signal decays with time because of the inhomogeneous field distribution of the flux-line lattice. The depolarization above \TC\ is reduced but still persists due to randomly orientated nuclear magnetic moments. The TF spectra were fit using a sinusoidal oscillating function with a Gaussian relaxation, an oscillatory background term arising from the muons implanted directly into the silver sample holder that do not depolarize, and a linear background.
\begin{multline}
G\left(t\right)=A_{1}\exp\left(-\frac{\sigma^2t^{2}}{2}\right)\cos\left(2\pi v_{1}t+\phi\right)\\ 
+A_{2}\cos\left(2\pi v_{2}t+\phi\right) + A_0,
\label{TF1}
\end{multline}
\noindent where $v_{1}$ and $v_{2}$ are the frequencies of the muon precession signal and the background signal respectively, $A_{0}$ is a flat background, and $\sigma$ is the Gaussian muon-spin relaxation rate. The superconducting contribution to the relaxation rate, $\sigma_{sc}$, can be calculated from $\sigma_{\mathrm{sc}}=\sqrt{\left(\sigma^2-\sigma_{\mathrm{n}}^2\right)}$ where $\sigma_{\mathrm{n}}$ is the nuclear magnetic dipolar contribution, which is assumed to be temperature independent.

For a type-II superconductor with a large upper critical field and a hexagonal Abrikosov vortex lattice, $\sigma_{sc}$ is related to the magnetic penetration depth $\lambda$ by the expression ${\frac{\sigma^{2}_{\mathrm{sc}}\left(T\right)}{\gamma^{2}_{\mu}} = 0.00371 \frac{\Phi^{2}_{0}}{\lambda^{4}\left(T\right) }}$ where ${\gamma_{\mu}/2\pi = 135.5~\mathrm{MHz/T}}$ is the muon gyromagnetic ratio and ${\Phi_{0} = 2.068 \times 10^{-15}~\mathrm{Wb}}$ is the magnetic-flux quantum. $\lambda$ is directly related to the superfluid density from which the nature of the superconducting gap can be determined. In the clean limit, the temperature dependence of the London magnetic penetration depth, $\lambda\left(T\right)$, can be calculated using the following expression:

\begin{multline}
\label{London}
\left[\frac{\lambda^{-2}\left(T,\Delta_{0,i}\right)}{\lambda^{-2}\left(0,\Delta_{0,i}\right)}\right]=1+ \\
\frac{1}{\pi}\int^{2\pi}_{0}\int^{\infty}_{\Delta_{\left(T,\phi\right)}}\left(\frac{\partial f}{\partial E}\right)\frac{EdE~d\phi}{\sqrt{E^2-\Delta_{i}\left(T,\phi\right)^2}},~
\end{multline}
\noindent where $f=\left[1+\exp\left(E/k_BT\right)\right]^{-1}$ is the Fermi function, and the temperature and angular dependence of the gap is $\Delta\left(T,\phi\right)=\Delta_{0}\delta\left(T/T_c\right)g\left(\phi\right)$. Here $g\left(\phi\right)$ is the angular dependence of the superconducting gap function and is 1 for an $s$-wave gap and $\lvert \cos\left(2\phi\right)\rvert$ for a $d$-wave gap where $\phi$ is the azimuthal angle along the Fermi surface. $\Delta\left(0\right)$ is the gap magnitude at zero kelvin and the temperature dependence of the gap is approximated by~\cite{Carrington} $\delta\left(T/T_c\right)=\tanh\left\{1.82\left[1.018\left(T_c/T-1\right)\right]^{0.51}\right\}$. For the multigap analysis we have used a weighted sum of the two gaps given by:
\begin{multline}
\label{weighting}
\left[\frac{\lambda^{-2}\left(T,\Delta_{0}\right)}{\lambda^{-2}\left(0,\Delta_{0}\right)}\right] = \\
w \left[\frac{\lambda^{-2}\left(T,\Delta_{0,1}\right)}{\lambda^{-2}\left(0,\Delta_{0,1}\right)}\right] + \left(1-w\right)\left[\frac{\lambda^{-2}\left(T,\Delta_{0,2}\right)}{\lambda^{-2}\left(0,\Delta_{0,2}\right)}\right].
\end{multline}
Figure~\ref{ZF_TFdata}(c) shows $\lambda^{-2}\left(T\right)$. We obtain good fits to the data, as measured by $\chi^2_{\mathrm{norm}}$ and the form of the normalized residual, using a two-gap $\left(s+s\right)$-wave model (see Fig.~\ref{ZF_TFdata}(c)). The $\lambda^{-2}\left(T\right)$ data were also fit using a single isotropic $s$-wave, a $d$-wave, and an $\left(s+d\right)$-wave model. The parameters extracted from these fits are given in Table~\ref{table_of_gapparameters}. There is little difference between the quality of the fits for the $\left(s+s\right)$ and $\left(s+d\right)$ models, as measured by $\chi^2_{\mathrm{norm}}$, with the $\left(s+s\right)$-wave model just preferred. Fig.~\ref{ZF_TFdata}(c) clearly shows that a single-gap $s$-wave model does not produce a good fit. The value of the larger energy gap for the $\left(s+s\right)$-wave model is $\Delta_{0,1} = 1.16(4)$~meV giving a superconducting gap ratio $\Delta_{0,1}/k_{\mathrm{B}}T_{\mathrm{c}} = 2.28(8)$, which is higher than the 1.76 expected for BCS superconductors; a further indication of the enhanced electron-phonon coupling in the superconducting state of \TRB. The $\left(s+s\right)$-wave model gives a magnetic penetration depth $\lambda(0) = 648(5)$~nm, which is higher than the value calculated from magnetization~\cite{SupplMat, Carnicom}. 

\begin{table}
\caption{Superconducting gap parameters for \TRB\ extracted from the fits to the penetration depth data using a BCS model in the clean limit.}
\label{table_of_gapparameters}
\begin{center}
\begin{tabular}[t]{lccc}\hline\hline
Model ~~& $d$-wave~~~& $\left(s+s\right)$-wave ~~~& $\left(s+d\right)$-wave \\\hline
w & 1 & $0.89(2)$~& $0.51(2)$ \\
$\Delta_{0,1}$~(meV)& $1.92(7)$~& $1.16(4)$ & $1.30(6)$  \\
$\Delta_{0,2}$~(meV)& - & $0.14(5)$~& $1.44(5)$ \\
$\Delta_{0,1}/k_{\mathrm{B}}T_{\mathrm{c}}$~& $3.78(14)$ & $2.28(8)$ & $2.56(11)$ \\
$\Delta_{0,2})/k_{\mathrm{B}}T_{\mathrm{c}}$~& - & $0.27(9)$ & $2.83(9)$ \\
$\lambda^{-2}\left(0\right)$~($\mu$m$^{-2}$) & $2.44(2)$ & $2.38(2)$ & $2.37(4)$ \\
$\chi^2_{\mathrm{norm}}$  & 1.81 & 1.18 & 1.35 \\
\hline\hline
\end{tabular}
\end{center}
\end{table}

\textit{Discussion}:- Muon spectroscopy, heat capacity, and resistivity measurements have been carried out on the new chiral NCS superconductor \TRB. We show \TRB\ is a bulk type II superconductor with a $T_{\mathrm{c}}=6.00(5)$~K driven by moderate electron-phonon coupling. Zero-field $\mu$SR shows time-reversal symmetry is preserved in the superconducting state. Our results clearly show strong evidence for multigap superconductivity in  \TRB. The temperature dependence of both the electronic heat capacity and the penetration depth, extracted from TF $\mu$SR data, reveal this multigap nature, with either an isotropic $\left(s+s\right)$-wave or an $\left(s+d\right)$-wave order parameter. For both models, the value of the larger gap $\Delta_{0,1}/k_{\mathrm{B}}T_{\mathrm{c}}$ extracted from the $\mu$SR and heat capacity data is higher than the BCS value of 1.76. 

We determine a number of other important superconducting parameters for this new NCS superconductor with a chiral structure~\cite{SupplMat}. In particular, we show that the upper critical field $\mu_0H_{\mathrm{c2}}\left(0\right)$ exceeds the Pauli limit. There are several potential sources for the suppression of paramagnetic pair breaking. Any triplet superconducting component should be robust to paramagnetic pair breaking~\cite{Bauer12, Smidman17}, but \TRB\ appears to be a superconductor in the dirty limit with considerable localization of the carriers and perhaps some disorder due to variations in the stoichiometry~\cite{SupplMat}. This is likely to suppress any triplet component that may be present. Strong-coupling can increase $H_{\mathrm{c2}}\left(0\right)$. The fit to the $C_{\mathrm{el}}\left(T\right)$ below \TC\ as well the jump in heat capacity at \TC\ suggest the coupling is only moderately enhanced, however, multigap behavior can sometimes reduce the jump at $T_{\mathrm{c}}$~\cite{BiswasLu2fe3Si5, Zehetmayer}. Spin-orbit scattering associated with the presence of heavier elements Ta and Rh could help to increase $H_{\mathrm{c2}}\left(0\right)$ in \TRB\ (cf. Ta$_2$Pd$_x$S$_5$ and related $R_2$Pd$_x$S$_5$ materials~\cite{Zhang13,Lu14}). $H_{\mathrm{c2}}\left(0\right)$ can also be increased by a Stoner enhancement~\cite{Fuchs08}.

The observed multigap nature of the superconductivity is also likely to play an important role in augmenting the upper critical field~\cite{Gurevich}. Further investigation into the effects driving this enhancement in the upper critical field are necessary, but this will require, for example, a much more detailed understanding of the density of states. High-quality single crystals are urgently required to further investigate the superconducting order parameter of \TRB\ and to confirm the mechanism allowing for a Pauli-limit violation in this compound.

\begin{acknowledgments}
This work is funded by the EPSRC, United Kingdom, through grant EP/M028771/1. This project has received funding from the European Research Council (ERC) under the European Union’s Horizon 2020 research and innovation programme (Grant agreement No. 681260).
\end{acknowledgments}

\bibliography{TaRh2B2_DM_References}

\end{document}